# Successive Phase Transitions in a Metal-Ordered Manganite Perovskite $YBaMn_2O_6$

T. Nakajima, H. Kageyama, and Y. Ueda

*Institute for Solid State Physics , University of Tokyo, Kashiwanoha, Kashiwa, Chiba, 277-8581, Japan*

## Abstract

Structural, magnetic and electric properties of a metal-ordered perovskite $YBaMn_2O_6$ have been studied by means of powder X-ray diffraction, DSC, magnetic susceptibility, and electric resistivity. It is found that this material undergoes a 1st order structural phase transition at $T_{c1}$=520 K, from a pseudo orthorhombic (monoclinic) phase ($T_{c1}<T$) to a pseudo tetragonal (monoclinic) one ($T<T_{c1}$). Accompanied by this transition, the susceptibility exhibits a sharp drop, where ferromagnetic exchange interaction becomes antiferromagnetic. Furthermore, two more transitions, a metal-insulator transition at $T_{c2}$=480 K and an antiferromagnetic ordering at $T_{c3}$=200 K, have been observed respectively.

## 1. Introduction

Recently, magnetic and electric properties of manganite perovskite with a general formula $(A^{3+}_{1-x}A'^{2+}_{x})MnO_3$, where $A^{3+}$ and $A'^{2+}$ represent rare earth and alkaline earth ions, respectively, have been extensively investigated. Among the interesting features are the so-called colossal magnetoresistance observed in, e.g., $La_{0.7}Sr_{0.3}MnO_3$ [1] and metal-insulator (M-I) transition accompanied by charge and orbital ordering observed in, e.g., $Pr_{0.5}Sr_{0.5}MnO_3$ [2]. It is now widely accepted that these enchanting phenomena are caused by the strong correlation/competition of multi-degrees of freedom, that is, spin, charge, orbital and lattice itself [3,4].

Unfortunately enough, almost all the works devoted to the series of manganite perovskite so far are on the disordered perovskite with the $A^{3+}$ and $A'^{2+}$ ions being randomly distributed. This means that, whenever $x$ is finite, there inevitably exists a disorder in the lattice. Since the physical properties of the manganite perovskite are quite sensitive to even a tiny change in lattice distortion [5], it is desirable to employ "cleaner" compound in order to understand the effect of A-site disorder more systematically.

When $x$ equals to 0.5, one can overcome this problem by preparing a metal-ordered double perovskite $AA'Mn_2O_6$, the structure of which consists of alternate stack of the $A^{3+}$ and $A'^{2+}$ ions along the $c$-axis as illustrated in Fig. 1. As far as the authors know, only the ordered $LaBaMn_2O_6$ was reported [6]. In this work, we successfully prepared the ordered $YBaMn_2O_6$ and investigated the relation between structural and physical properties of this material.

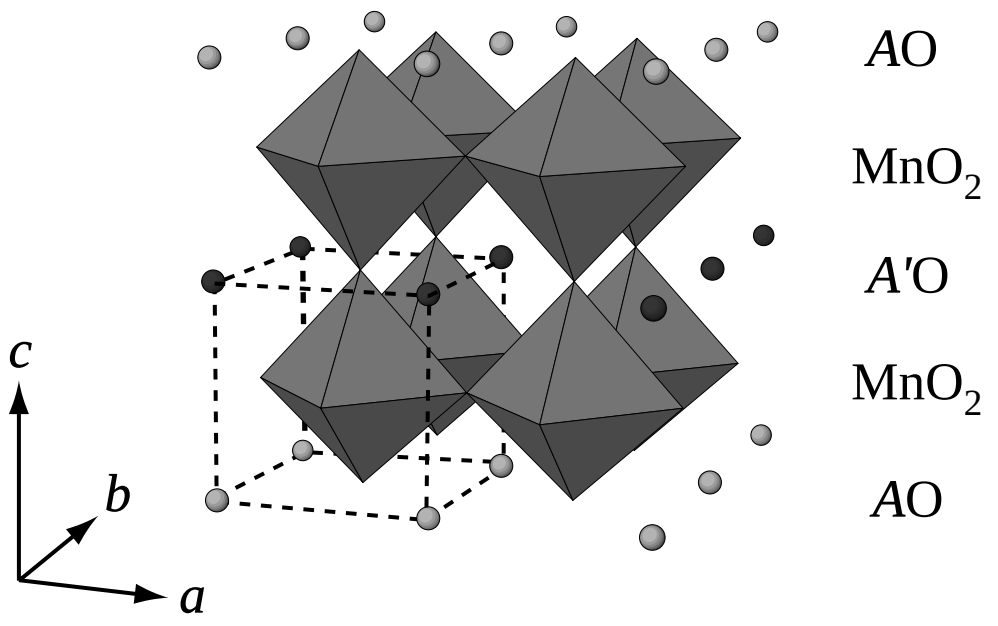

Fig. 1: Crystal structure of the ordered $AA'Mn_2O_6$. The broken lines represent the unit cell for the simple perovskite ($a_p \times b_p \times c_p$).

## 2. Experimental

The synthesis of the ordered perovskite requires several complex processes as described in Refs. [7] and [8]. Starting materials $Y_2O_3$, $MnO_2$, and $BaCO_3$ were ground thoroughly, pressed into pellets and calcined in an Ar flow at 1300 °C for 48 h which yields the ordered $YBaMn_2O_{6-y}$ ($y$~0.9) with a large amount of oxygen vacancy in the YO layer. Then, annealing it in flowing $O_2$ at 500 °C for 48 h leads to the fully oxidation of the sample, i.e., to $y \rightarrow 0$.

Thermogravimetric reduction in $H_2$ confirmed the oxygen content of the sample to be ideal stoichiometry $O_{6.0}$. The crystal structures were determined for 300-573 K by powder X-ray diffraction using $CuK_\alpha$ radiation. Differential scanning calorimetric (DSC) measurements was carried out for 150-600 K. The heating/cooling rate was typically 5-20 K/min.

The magnetic properties were studied using a SQUID magnetometer in a temperature range $T$ =5-600 K in a magnetic field of 1000 Oe. The electric resistivity of a sintered pellet was measured for $T$ =100-620 K by a conventional four-probe technique.

### 3. Results and Discussion

The X-ray diffraction profile scanned at room temperature (Fig. 2 (a)) indicates that the obtained sample is a mixture of a main phase of the ordered $YBaMn_2O_6$ and a miner impurity phase of $BaMnO_{3-\delta}$ (about 3%). The (001) reflection around $2\theta$ =12° assures the successful alternation of $Y^{3+}$ and $Ba^{2+}$ cations along the $c$-axis, thus doubling the $c$ parameter with respect to the simple perovskite. It appears that the crystal structure is tetragonal, but the careful analysis of the data finally led to the conclusion that it is monoclinic with the unit cell $a_p \times b_p \times 2c_p$, where $a_p$, $b_p$, and $c_p$ represent lattice parameters for the simple perovskite structure. The structural parameters are determined as follows: $a$=3.901(9) Å, $b$=3.898(1) Å, $c$=7.602(2) Å, $\beta$=90.2(1) °, and $V$=115.5(9) Å$^3$.

Figure 3 shows the temperature dependence of the structural parameters, where a clear jump is observed at $T_{c1}$=520 K, suggesting a 1st order transition. The nature of the 1st order phase transition at $T_{c1}$ is also visible in the DSC curve (Fig. 4), where a huge endothermic peak is observed on heating. We found that the structure of the high-$T$ phase is also consistent with the monoclinic symmetry with the unit cell of $a_p \times b_p \times 2c_p$. The structural parameters, for example, at 573K are determined as follows: $a$=3.930(6) Å, $b$=3.846(3) Å, $c$=7.731(2) Å, $\beta$=90.2(8) °, and $V$=116.9(8) Å$^3$. It is strange that the high-$T$ phase with a pseudo orthorhombic structure ($a$>$b$, $\beta \approx 90^o$) has apparently a lower symmetry than the low-$T$ phase with a pseudo tetragonal structure ($a \approx b$, $\beta \approx 90^o$). We suppose the low-$T$ phase may have superstructures originating from the charge/orbital ordering.

The careful reading of the DSC curve in Fig. 4 suggests that, besides the already-mentioned peak at $T_{c1}$, there are two weak endothermic peaks at $T_{c2}$=480 K and $T_{c3}$=200 K. In Fig. 5, we demonstrate the electric resistivity as a function of $T$, which can be characterized by the M-I transition that occurs at $T_{c2}$ and by small anomalies at $T_{c1}$ and $T_{c3}$. Although for $T_{c2}$<$T$ decreases with increasing $T$, it should be essentially metallic considering the relatively low and constant reisitivity (~$10^{-1}$ Ω·cm) and the fact that we used polycrystalline sample not a single crystal.

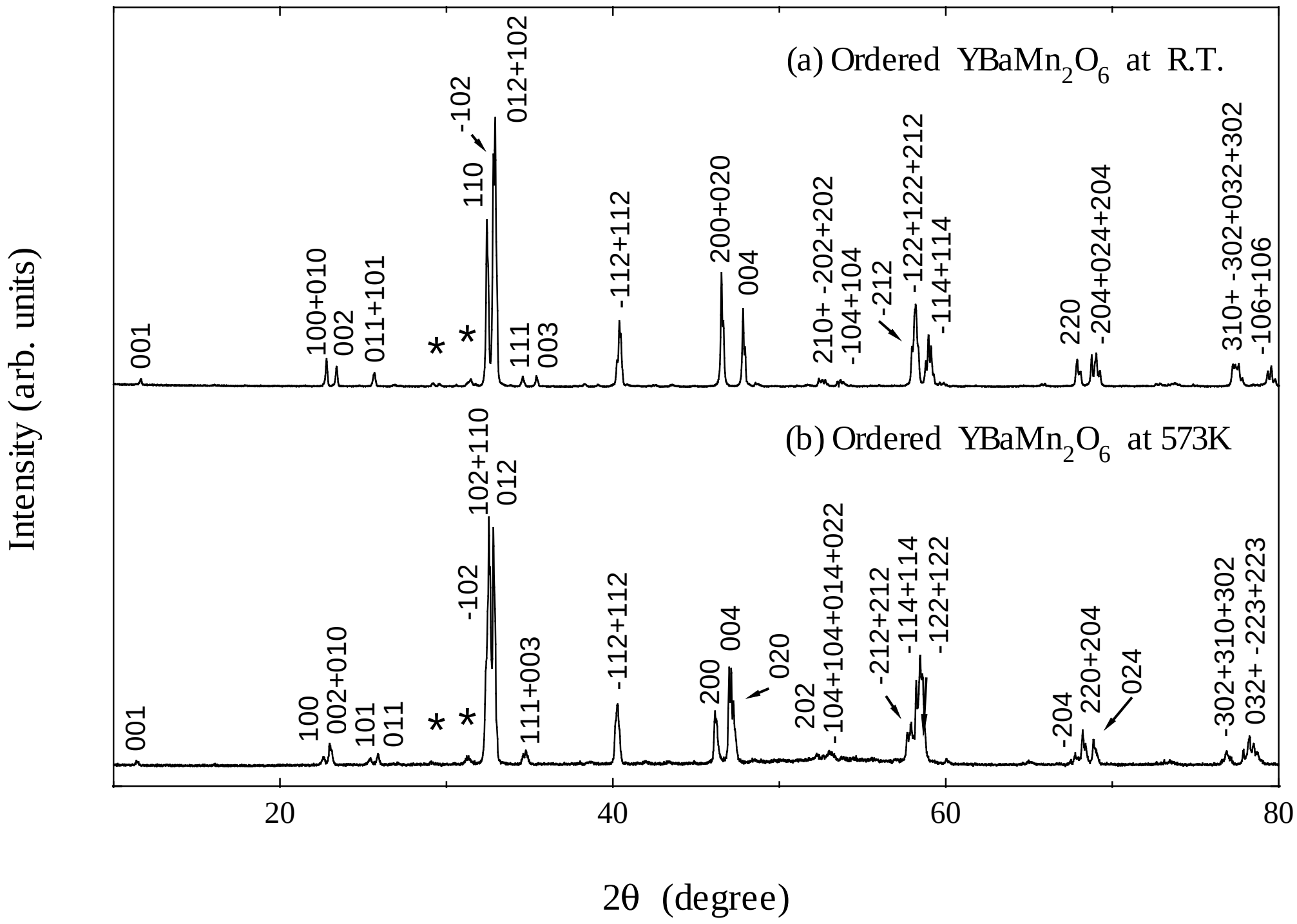

Fig. 2: X-ray diffraction data for the ordered $YBaMn_2O_6$ at (a) R.T. and (b) 573 K. Both profiles can be indexed by assuming a monoclinic structure with the $a_p \times b_p \times 2c_p$ unit cell. Peaks marked by asterisks come from an impurity phase $BaMnO_{3-\delta}$ and $Y_2O_3$.

Also shown in Fig. 5 is the temperature dependence of the magnetic susceptibility $\chi$ measured at 1000 Oe. At high temperature above $T_{c1}$, the $\chi$-$T$ curve shows a paramagnetic behaviour, obeying the Curie-Weiss law with the effective moment $P_{eff}$=7.12 $\mu_B$ and the Weiss constant $\theta_1$=286 K. The positive value of $\theta_1$ suggests the ferromagnetic spin correlation, and the value of $P_{eff}$ is compatible with a theoretical value of 7 $\mu_B$ derived from [$Mn^{3+}$ + $Mn^{4+}$] in $YBaMn_2O_6$. Upon decreasing $T$, drops sharply at $T_{c1}$, while only small anomaly is observed at $T_{c2}$ where the M-I transition occurs. Below $T_{c2}$, the $\chi$-$T$ curve once again shows a paramagnetic behaviour but the Curie-Weiss fitting gave a negative value of $\theta_2$= -379 K, indicating an antiferromagnetic interaction. Finally, exhibits a steep drop at $T_{c3}$, then reaching a constant value ~0.006 emu/mol Mn. Considering the negative value of $\theta_2$, it is natural to assume that the transition at $T_{c3}$ would be due to the antiferromagnetic long-range order.

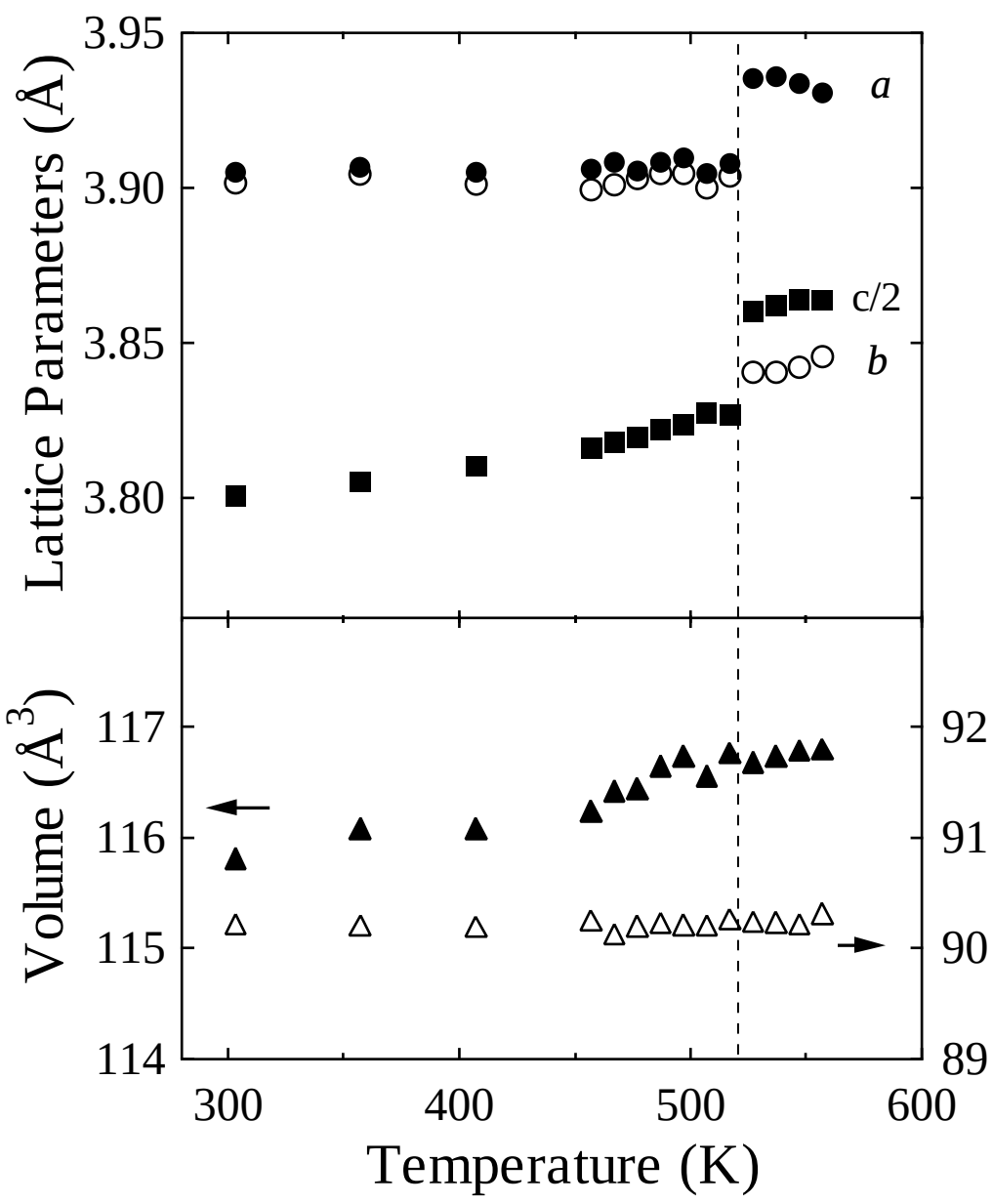

Fig. 3: Temperature variation of the reduced cell parameters $a$, $b$, $c$/2, $\beta$, and $V$ for the ordered $YBaMn_2O_6$.

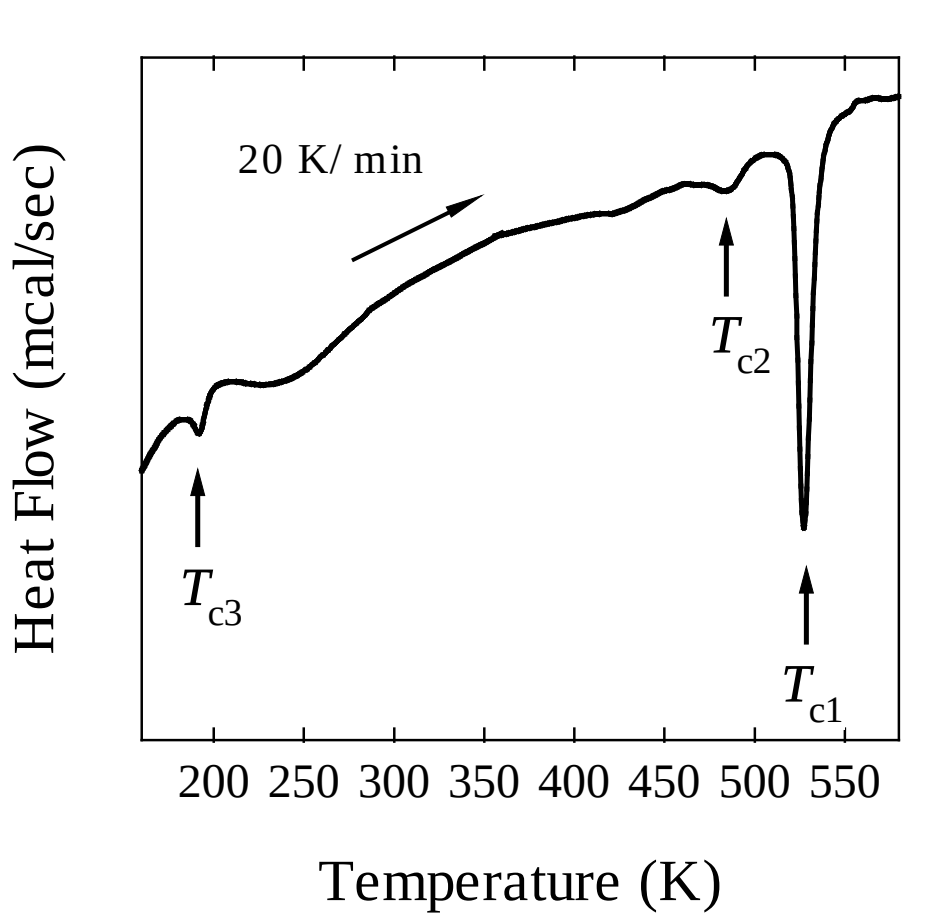

Fig. 4: DSC curve of the ordered $YBaMn_2O_6$ upon heating with a speed of 20 K/min.

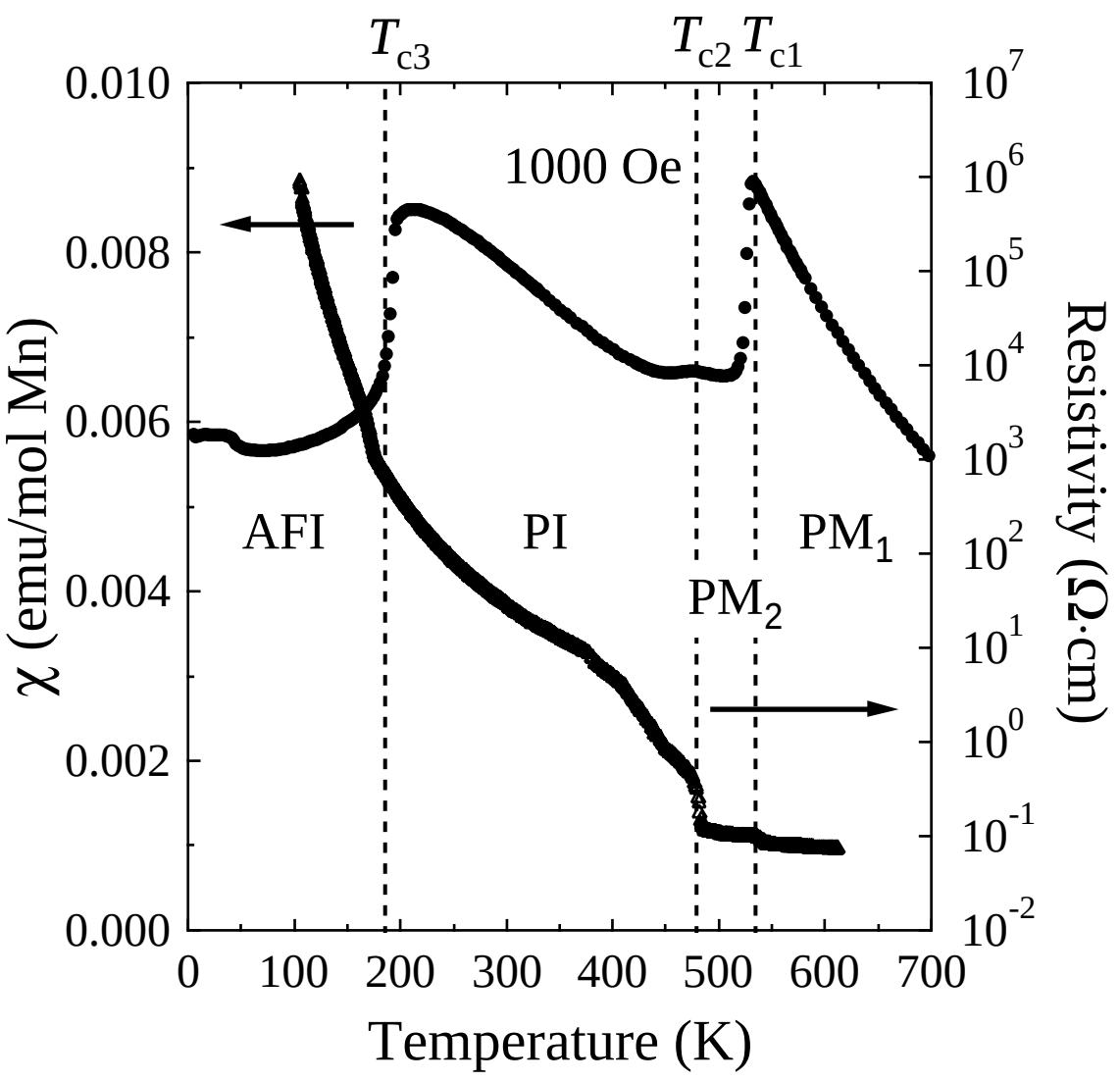

Fig. 5: Magnetic susceptibility and electric resistivity of the ordered $YBaMn_2O_6$ as a function of temperature.

As mentioned above, the ordered $YBaMn_2O_6$ shows three transitions, namely, from the paramagnetic metal with the nearly orthorhombic structure ($PM_1$) to the paramagnetic metal with the nearly tetragonal structure ($PM_2$) at $T_{c1}$, then to the paramagnetic insulator (PI) at $T_{c2}$, and finally to the antiferromagnetic insulator (AFI) at $T_{c3}$. Now let us discuss the successive phase transitions from the viewpoint of multi-degrees of freedom. The $PM_1$ state with the positive $\theta$ implies that the dominant ferromagnetic interactions at high temperatures originate from the double-exchange mechanism. Accordingly, without the transitions at $T_{c1}$ and $T_{c2}$, this compound would experience the ferromagnetic ordering. What we would like to emphasize is that in $YBaMn_2O_6$ the charge/orbital ordering occurs at much higher temperature compared with the disordered perovskite. Take $Sm_{0.5}Ca_{0.5}MnO_3$ [9] and $Pr_{0.5}Sr_{0.5}MnO_3$ [2], for example, the charge/orbital ordering temperature is reported to be, respectively, 270 K and 140 K. This suggests that the disorder in the lattice originating from the random distribution of $A$ and $A$' cations significantly lowers the charge/orbital ordering temperature.

It has been known that the charge/orbital ordering is usually accompanied by the structural phase transitions as is seen in Ref. [10]. In the present compound, however, charge/orbital ordering temperature ($T_{c2}$) is lower than the structural phase transition ($T_{c1}$). This is significantly different from the transitions reported in the disordered manganite perovskite so far.

To fully understand the nature of the successive phase transition in particular intermediate phase ($PM_2$), it is necessary to perform microscopic measurements such as neutron diffraction and NMR, which are now in progress.

## 4. Conclusion

In summary, we studied the structural and physical properties of the ordered manganite $YBaMn_2O_6$ to find three phase transitions at $T_{c1}$=520 K ($PM_1 \rightarrow PM_2$), at $T_{c2}$=480 K ($PM_2 \rightarrow$ PI) and at $T_{c3}$=200 K (PI $\rightarrow$ AFI). At $T_{c1}$, the structure changes from pseudo orthorhombic to pseudo tetragonal one. The M-I transition which occurs at $T_{c2}$ would be accompanied by the charge/orbital ordering. Reflecting the ordering between $Y^{3+}$ and $Ba^{2+}$ ions, the transition temperature is much higher than that reported for disordered manganite. The antiferromagnetic ordering occurs at $T_{c3}$.

## Acknowledgements

This study was partly supported by Grant-in-Aid for Scientific Research and for Creative Scientific Research from the Ministry of Education, Science, Sports and Culture.